\documentclass[copyright,creativecommons]{eptcs}
\providecommand{\event}{WWv2012} 
\usepackage{breakurl}             

\usepackage{rotating}
\usepackage{pdflscape}
\usepackage{graphicx}
\usepackage{graphicx}
\usepackage{flushend}
\usepackage{graphicx}
\usepackage{colortbl}
\usepackage{xcolor}
\usepackage{colortbl}
 
\graphicspath{{../pdf/}{../jpeg/}}
\DeclareGraphicsExtensions{.pdf,.jpeg,.png}
\usepackage{url}

\definecolor{light-gray}{gray}{0.95}  
  
\usepackage{listings}
\usepackage{times,amsmath,epsfig}

\usepackage{url}
 
\title{Model Validation in Ontology Based Transformations\thanks{This work has been supported by the Spanish Ministry MICINN and Ingenieros Alborada IDI under grant TRA2009-0309. This work has been also supported by the EU (FEDER) and the Spanish Ministry MICINN under grants TIN2010-15588, TIN2008-06622-C03-03, and the JUNTA ANDALUCIA (proyecto de excelencia) ref. TIC-6114.
}
}
\author{Jes\'us M. Almendros-Jim\'enez
\email{jalmen@ual.es}
\institute{Dpto. de Lenguajes y Computaci\'on}
\institute{Universidad de Almer\'\i a\\ 04120-Spain}\\
\and 
Luis Iribarne
\email{luis.iribarne@ual.es}
\institute{Dpto. de Lenguajes y Computaci\'on}
\institute{Universidad de Almer\'\i a\\ 04120-Spain}\\
}
\def\titlerunning{Model Validation in Ontology Based Transformations}
\def\authorrunning{Jes\'us M. Almendros-Jim\'enez \& Luis Iribarne}
\begin{document}
\maketitle

\begin{abstract}
Model Driven Engineering (MDE) is an emerging approach of software engineering. 
MDE emphasizes the construction of models from which the implementation should be
derived by applying model transformations. 
The Ontology Definition Meta-model (ODM) has been proposed as a profile for UML models of the 
Web Ontology Language (OWL).  In this context, transformations of UML models can be mapped into ODM/OWL 
transformations. On the other hand, model validation is a crucial task in model transformation. 
Meta-modeling permits to give a syntactic structure to source and target models. However,
semantic requirements have to be imposed on source and target models. A given transformation
will be sound when source and target models fulfill the syntactic and semantic requirements.
In this paper, we present an approach for model validation in ODM based transformations.
Adopting a logic programming based transformational approach we will show how it is possible
to transform and validate models. Properties to be validated range from structural and semantic
requirements of models (pre and post conditions) to properties of the transformation (invariants).
The approach has been applied to a well-known example of model transformation:
the Entity-Relationship (ER) to Relational Model (RM) transformation.
\end{abstract}

\section{Introduction}
 
 \emph{Model Driven Engineering (MDE)} is an emerging approach for software development.
MDE emphasizes the construction of models from which the implementation should be
derived by applying model transformations. Hence, the
\emph{model transformation} \cite{Tratt,Interoperability}
is a key tool of MDE.  According to the \emph{Model Driven Architecture (MDA)} \cite{MDA} initiative of the
\emph{Object Management Group (OMG)} \cite{OMG}, the model transformation
provides to developers a framework for transforming their models. 

The MDA approach proposes (at least) three levels
in order to describe a model transformation: the first one is the so-called \emph{meta-meta-model}, which is the basis of
the model transformation, and provides the language for describing meta-models. The second one
consists in the \emph{meta-models} of the models to be transformed. Source and target models must conform
to the corresponding meta-model. Such meta-models are modeled according to the meta-meta-model.
The third one consists in the source and target \emph{models}. Source and target models are instances of the corresponding 
meta-models. In addition, source and target meta-models are instances of the meta-meta-model.
In order to define a model transformation one should be able to meta-model the source and target models with regard to the meta-meta-model, and map source and target meta-models. Model transformation needs formal techniques for specifying the transformation. In most cases transformations can be expressed
in some kind of \emph{rules}. 
   
On the other hand, the {\it Ontology Definition Metamodel (ODM)} proposal \cite{odm} of the {\it OMG} aims
to define an ontology-based representation of UML models. ODM is a standard for representing 
UML models by OWL in which, among others,
UML classes are mapped into ontology concepts, UML  associations are mapped into ontology roles, and
multiplicity restrictions of UML are mapped into cardinality restrictions in roles. ODM is itself an UML meta-model
in which UML models can be  accommodated.  Following the ODM proposal,
 an UML model can be represented by an ontology in which the {\bf TBox} (i.e. the \emph{terminological box}) contains the UML
 meta-model while the {\bf ABox} (i.e. the \emph{assertional box}) contains the instance of the UML meta-model 
 which represents the model.
 
Model validation is a key element of MDE. Firstly, (a) source and target models must conform to
the corresponding meta-models. Source and target meta-models describe the {\it syntactic structure} of source 
and target models. However, some {\it semantic requirements} have to be imposed on source and 
target models. In UML semantic
requirements are usually expressed in the {\it Object Constraint Language (OCL)} \cite{OCL}. Secondly, (b) 
{\it pre-conditions} and {\it post-conditions} and {\it invariants}
are imposed on transformations. While source and target models can be well-formed with regard to meta-models, some extra requirements can be required. We can distinguish two specific cases: (b.1) {\it source and target model requirements},
and (b.2) {\it transformation requirements}. The first case covers requirements of source and target models in isolation. 
The second case covers requirements on target models with regard to the source models.

In the ODM context, one can argue that the use of OWL as modeling language provides a suitable 
framework for validation of properties. OCL can be replaced by OWL when specifying requirements imposed on models.
{\it OWL reasoning} is a widely studied topic of research, and many tools
have been developed in this context (for instance, the {\it Prot\'eg\'e} tool \cite{Protege} and the OWL reasoners
{\it Hermit}, {\it Jena}, {\it Fact++}, {\it Racer}, among others). OWL reasoning ranges from {\it ontology consistence testing}
to {\it ontology-based inference} (i.e. derivation from ontology axioms). The topic can be applied to ODM (an hence to UML) however, validation in model transformation is a wider topic of research.
Model validation involves ontology consistence testing and ontology-based inference for cases (a) and (b.1),  
while case (b.2) involves cross validation of ontologies.  

On the other hand, the relationship between logic programming and ontologies is well-known. 
OWL is based on the {\it Description Logic (DL)}  \cite{DL}, 
a family of fragments of {\it first order logic}, and some DL fragments can be encoded
into logic programming, for instance, the so-called {\it Description Logic Programming} approach \cite{Grosof}, 
and OWL RL \cite{OWLRL,almendros2011prolog}. Typically, Description Logic is used for representing 
the {\bf TBox}  and the {\bf ABox}. 
 The encoding of (fragments of) DL into logic programming is based on the representation of
 the {\bf TBox} by Prolog rules and the representation of the {\bf ABox} by Prolog facts.
 It means that ontology instances are represented by Prolog facts.

In this paper, we present an approach for model validation in ODM based transformations.
Adopting a logic programming based transformational approach we will show how it is possible to
transform and validate models. Properties to be validated range from structural and semantic
requirements of models (pre and post conditions) to properties of the transformation (invariants).
The approach has been applied to the well-known example of model transformation:
the Entity-Relationship (ER) into Relational Model (RM) transformation.
We have validated our proposal  in a prototype developed under SWI-Prolog. The prototype together
with the case study can be downloaded from \url{http://indalog.ual.es/mdd}.

The proposal is based on the use of logic programming with two ends.
Firstly, specification of transformations. Secondly, specification of properties for model validation.
Our approach adopts the OWL to logic programming mapping as basis.
Firstly, transformations can be expressed in Prolog rules. 
In model transformation, a transformation maps the source model into the target model,
and thus, the {\bf ABox} of the source model into the {\bf ABox} of the target model.
Model transformation can be seen as a mapping of Prolog facts following
the OWL to logic programming encoding, and it can be defined by Prolog rules. 
Secondly, model validation can be encoded with Prolog.  Using Prolog atoms and some elements of 
 Prolog meta programming  we are able to validate source and target models as well as transformations.  

The advantages of the approach are the following. Firstly, the declarative nature of the specification, secondly,
the use of a standardized language (Prolog), and the ability of executing transformations and automatically
validate source and target models. Besides, the use of Prolog
as validation language enriches the mechanisms of ODM and OWL constructors for expressing model requirements.
While ODM is an OWL profile for UML models the expressivity power of ODM is limited,  
and model validation needs to express more complex requirements.

The structure of the paper is as follow. Section 2 will introduce the model transformation framework
and will describe a case study of transformation. Section 3 will present the Prolog-based approach.
Section 4 will show model validation. Section 5 will discuss related work. Finally, 
Section 6 will conclude and present future work.

\section{Model Transformation}

The elements to be considered in a ontology based transformation
using Prolog as transformation language can be summarized as follows:
\begin{itemize} 
\item[--] 
We have to consider the {\it meta-model of the source model} which defines the elements
occurring in the source model.  Instances of the source meta-model are transformed by applying the 
transformation rules. 
\item[--] 
We have to consider the {\it meta-model of the target model} which defines the elements 
occurring in the target model. 
\item[--] 
We have to define {\it Prolog rules} for transforming instances of the source meta-model into instances
of the target meta-model.  
\item[--] 
We have to define {\it Prolog rules} for validating the transformation. The validation consists in 
source model validation, target model validation and transformation validation. 
\end{itemize}

The question now is, how to express transformations and validations in Prolog? Our proposal is as follows. 
\begin{itemize}
\item[--]
The ODM proposal provides a representation of UML models by an ontology. 
The {\bf TBox} represents the meta-model and the {\bf ABox} properly represents the model.
We can represent the {\bf ABox} by Prolog facts. Fortunately, SWI-Prolog, used in our prototype,
is equipped with a library for RDF(S)/OWL which imports and exports RDF triples to Prolog facts.
\item[--] 
A transformation from a source model into a target model can be seen as a transformation
from the set of Prolog facts of the source model into the set of Prolog facts representing the target model. 
Prolog rules can be used for transforming Prolog facts.
\item[--] 
Model validation consists in checking properties on source and target models in isolation as well as
checking cross properties on both models. Model validation with Prolog consists in checking properties
about the set of Prolog facts representing source and target models.
\end{itemize}

\begin{figure*}[t!]
\begin{center}
\includegraphics*[width=12cm]{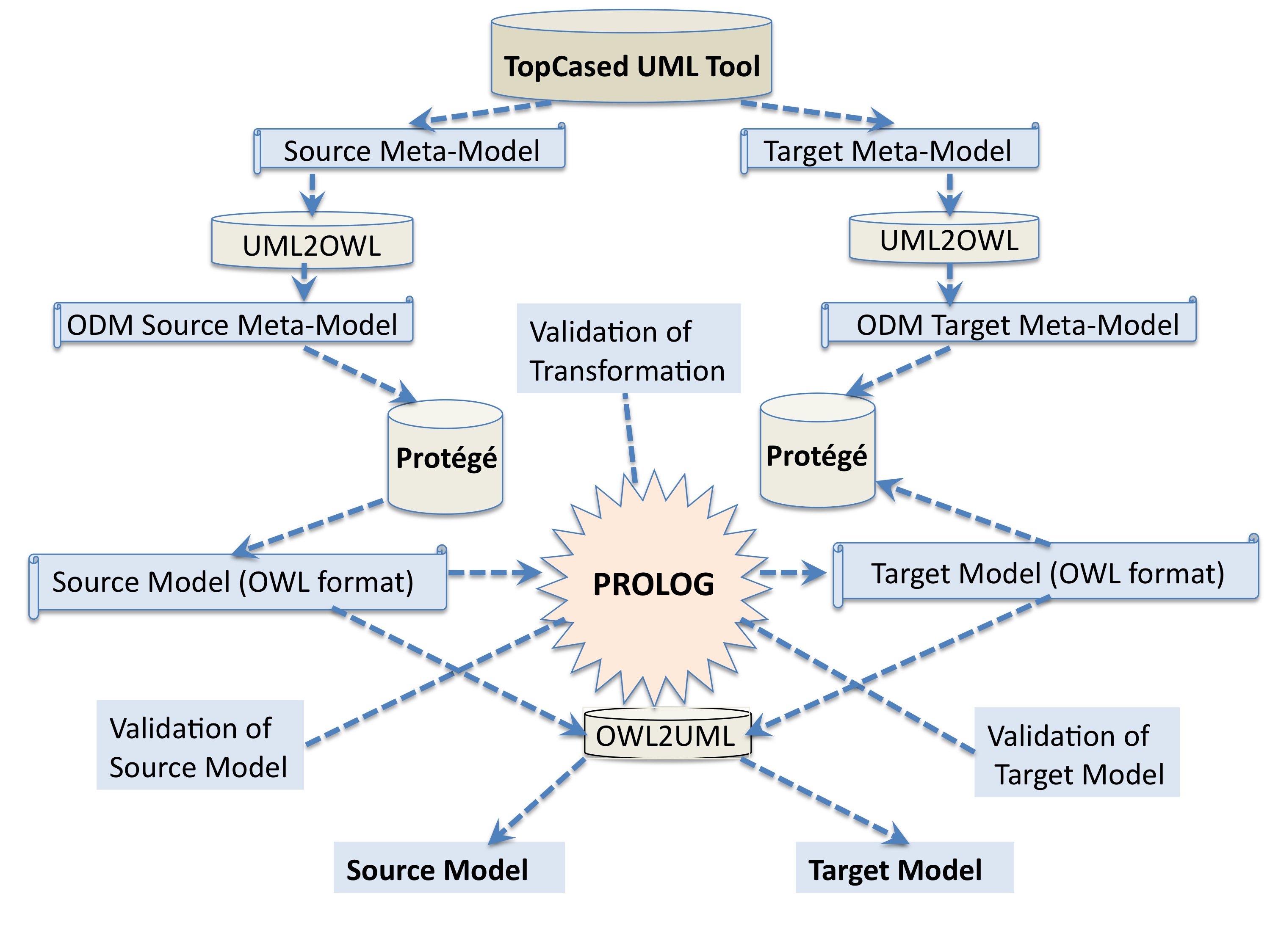}
\end{center}
\caption {UML/OWL tools} \label{architecture}
\end{figure*}

Our approach has been implemented and tested with some examples. We have 
used several UML and OWL tools (see Figure \ref{architecture}). 
\begin{itemize}
\item[--] We have used the {\it TopCased} UML tool \cite{TopCased} for designing the source and target meta-models. 
\item[--] In addition, we made use of a
 UML2OWL transformer (available from \cite{UML2OWL}) in order to have the ODM-based representation
 of source and target meta-models. 
\item[--] We have also employed the {\it Prot\'eg\'e} tool \cite{Protege}
 for defining the instance of the source meta-model, and for exporting the source model (i.e. meta-model+instance)
to an OWL document. 
\item[--] After, the SWI-Prolog interpreter is utilized 
to validate the source model, and
to transform the instance of the source model
into the instance of the target model. 
\item[--] Once the target model is computed, SWI-Prolog is used to validate 
the target model, and to validate the transformation. 
\item[--] Next, the {\it Prot\'eg\'e} tool is also employed to 
export the target model together with the target meta-model to an OWL document.
\item[--] Finally, an OWL2UML transformer has been employed to obtain the target model from the ODM-based representation.
\end{itemize}
 
\subsection{Case Study}\label{Case}

\begin{figure*}[t!]
\begin{center}
\includegraphics*[width=13cm]{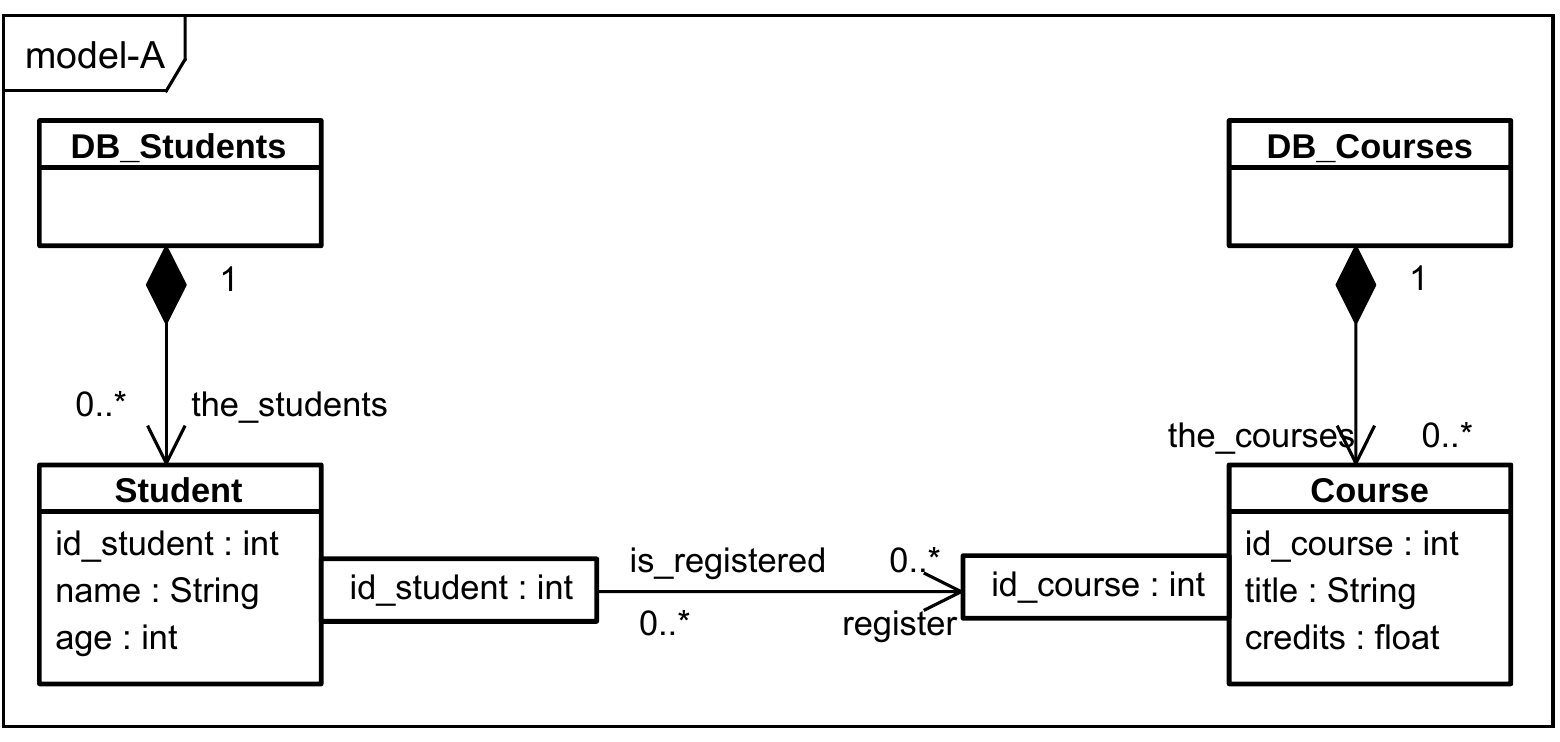}
\end{center}
\caption {Entity-relationship modeling of the Case Study} \label{fig:ModelA}
\end{figure*}

\begin{figure*}[t!]
\begin{center}
\includegraphics*[width=16cm]{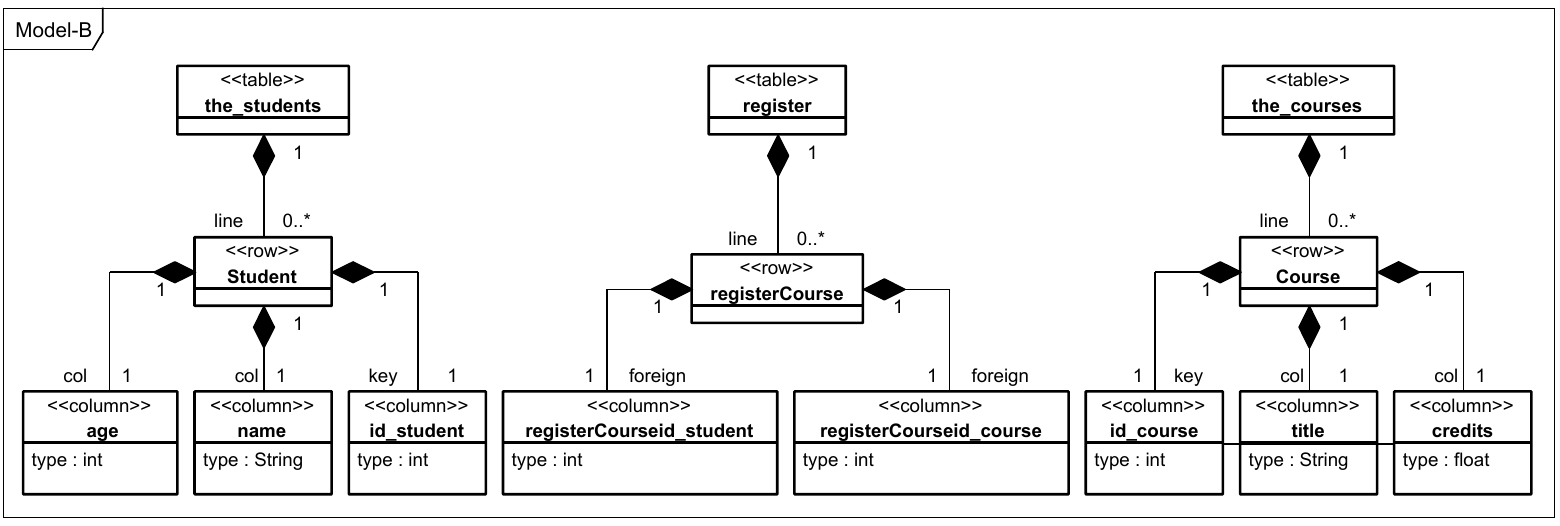}
\end{center}
\caption {Relational modeling of the Case Study} \label{fig:ModelB}
\end{figure*}

In this section we will describe the case study of the paper. It
is a well-known example of model transformation. Basically, the entity-relationship (ER) model
is transformed into the relational model (RM).  
The model of Figure \ref{fig:ModelA} represents the modeling of a database
by an ER style diagram, while the model of Figure  \ref{fig:ModelB} is a RM style modeling of the same database.

The ER modeling of Figure \ref{fig:ModelA} can be summarized as follows.
{\it Data} are represented by classes (i.e., {\it Student} and {\it Course}), including attributes;
{\it stores} are defined for each data  (i.e., {\it DB\_Students} and {\it DB\_Courses});  
{\it relations} are represented by associations;
relation names are association names; besides, association ends are defined (i.e., {\it the\_students}, {\it the\_courses}, {\it is\_registered} and {\it register}); relations can be adorned with {\it qualifiers} and navigability; 
 qualifiers specify the key attributes of each data (used as foreign keys of the corresponding association). 
 
\begin{figure*}[t!]

\begin{tabular}{ll}
\includegraphics*[width=8cm]{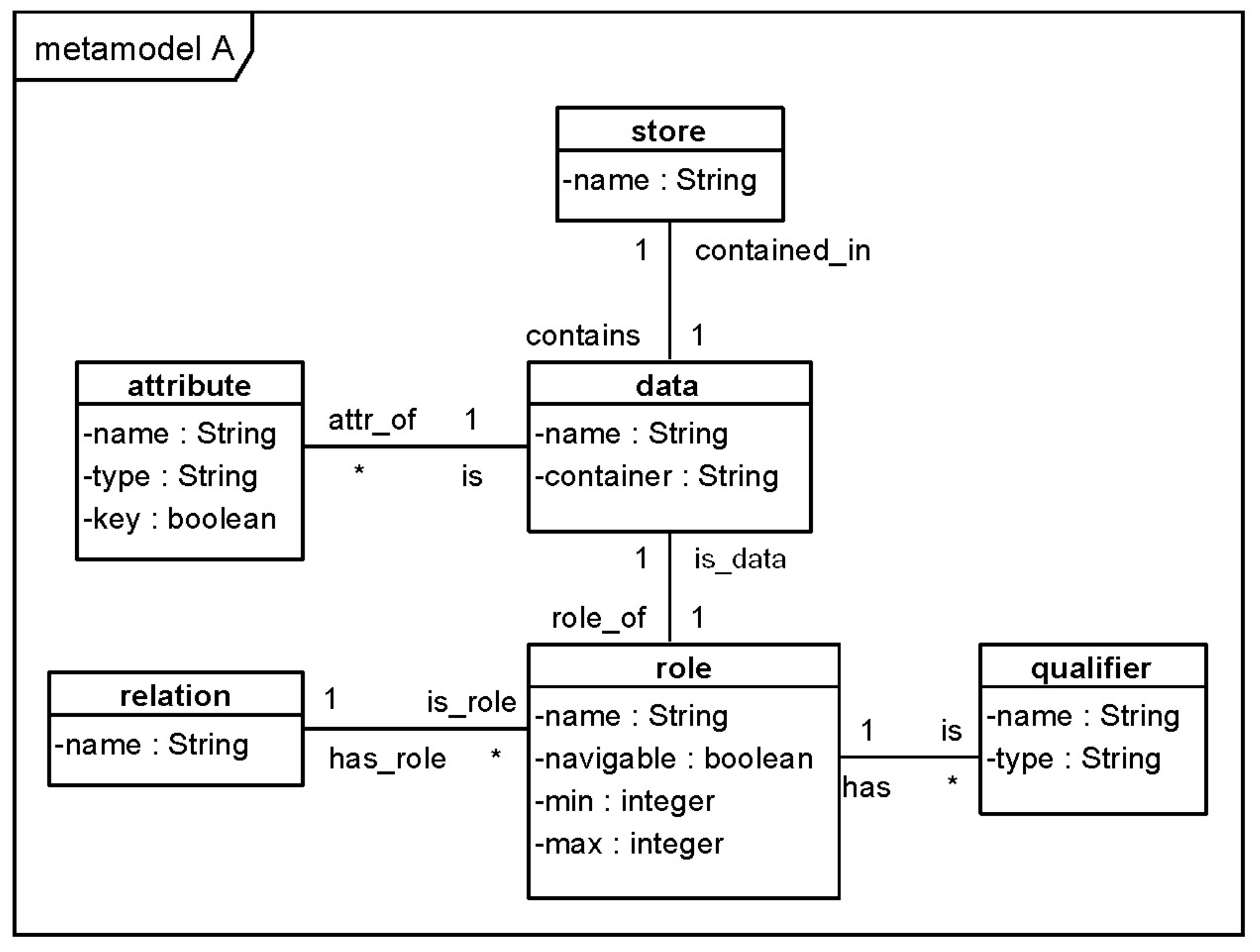} &
\includegraphics*[width=7cm]{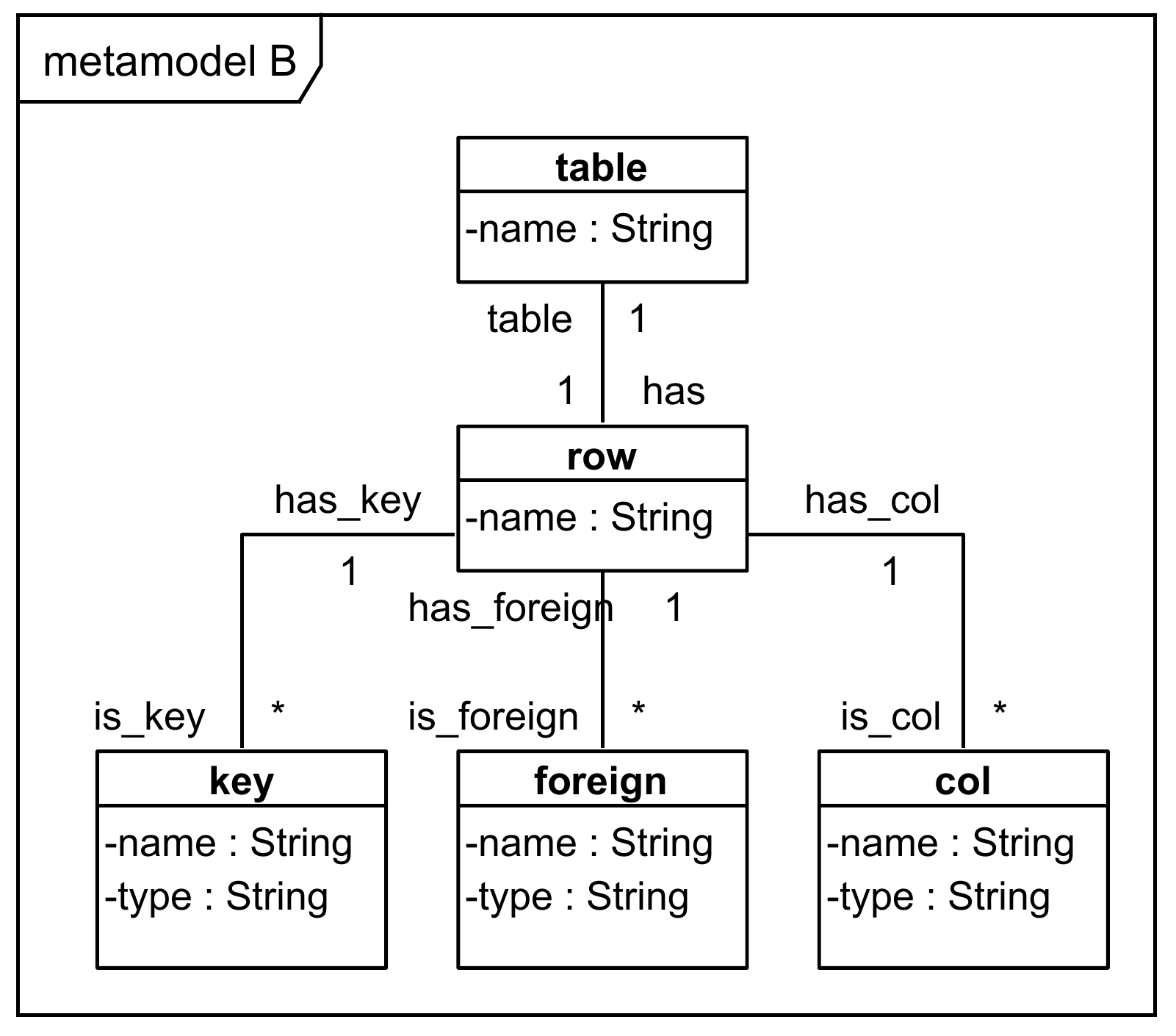}
\end{tabular}
\caption {Meta-model of the Source/Target Models} \label{fig:metamodel}
\end{figure*}
\begin{figure*}[!t]

\begin{center}
\includegraphics*[width=15cm]{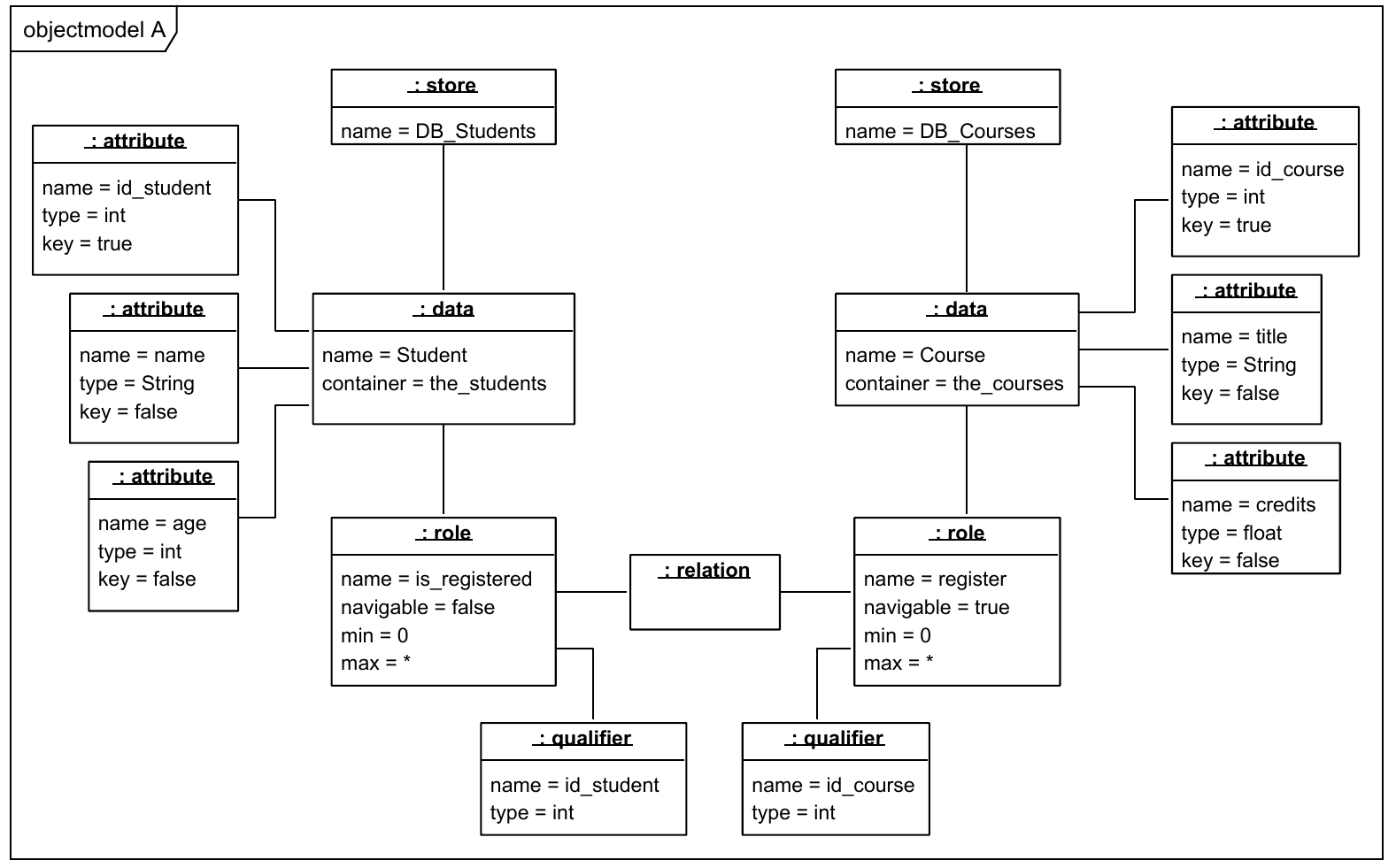}
\end{center}
\caption {Object Model of Source Model} \label{fig:metamodelObjectA}
\end{figure*} 
 \begin{figure*}[!t]

\begin{center}
\includegraphics*[width=16cm]{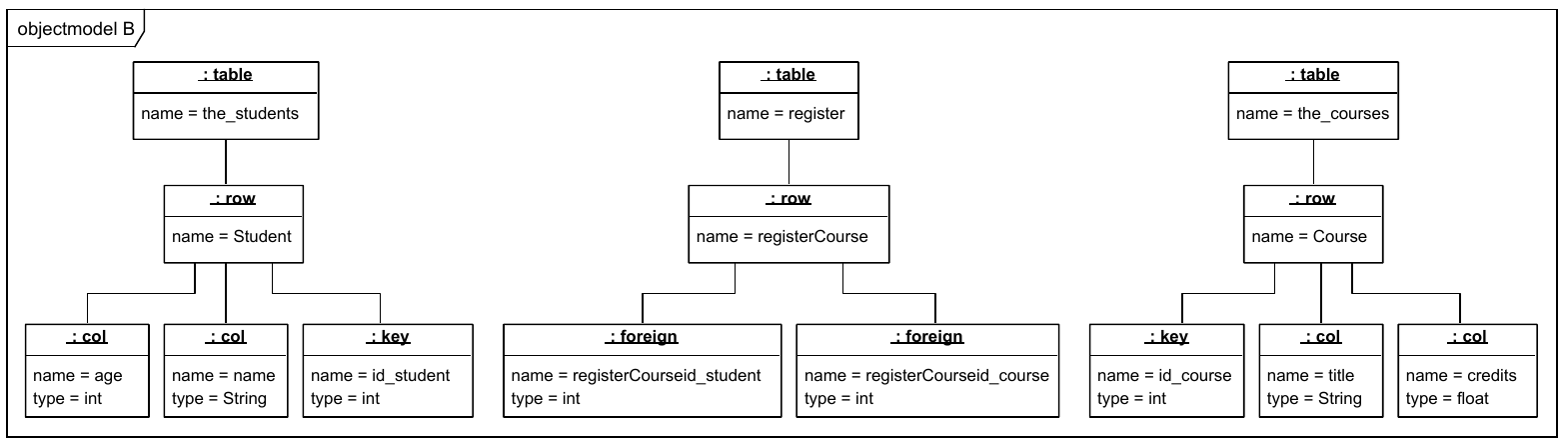}
\end{center}
 \caption {Object Model of Target Model} \label{fig:metamodelObjectB}
\end{figure*}

Figure  \ref{fig:ModelB} shows the RM modeling of the same
database.  {\it Tables} are composed of {\it rows}, and rows are composed of {\it columns}. 
It introduces the following UML stereotypes: {\it $<<$ table $>>$},
{\it $<<$ row $>>$} and {\it $<<$ column $>>$}. Furthermore, {\it line} is the role 
of the rows in the tables and {\it key}, {\it foreign} and {\it col} is the role of the key, foreign, and non key and non foreign
attributes in rows, respectively. Finally, each column
has an attribute {\it type}.
 
Figure  \ref{fig:metamodel}
represents the meta-models of ER and RM models.
In the first case, {\it DB\_Students} and {\it DB\_Courses} are instances of the class {\it store}, while 
{\it Student} and {\it Course} are instances of the class {\it data}, and the attributes of classes {\it Student} and {\it Course}
are instances of the class {\it attribute}. In the second case, tables and rows of the target
model are instances of the corresponding meta-classes, and the same can be said about classes {\it key}, {\it col} and {\it foreign}.
 
Now, the goal of the model transformation is to describe how to transform a class diagram
of the type A (like Figure \ref{fig:ModelA}) into a class diagram of the type B (like Figure \ref{fig:ModelB}).
 
The transformation is as follows. The transformation generates two tables
called {\it the\_students} and {\it the\_courses} each including three columns that are grouped into rows. The 
table {\it the\_students} includes for each student the attributes of {\it Student} of Figure \ref{fig:ModelA}.
The same can be said for the table {\it the\_courses}. Given that the association 
between {\it Student} and {\it Course} is navigable from {\it Student} to {\it Course}, 
a table of pairs is generated to represent the assignments of students
to courses, using the role name of the association end, that is, {\it register} concatenated with 
{\it Course}, that is, {\it registerCourse}, for naming the cited table. The columns {\it registerCourseid\_student} and {\it registerCourseid\_course} taken from qualifiers, 
play the role of foreign keys which are represented by the role {\it foreign} in the association
of Figure \ref{fig:ModelB}.

The transformation can be considered as a transformation between object diagrams of source and 
target meta-models (see Figures \ref{fig:metamodelObjectA} and \ref{fig:metamodelObjectB}). A transformation should be able to define a set of rules from which 
instances of the target meta-model are obtained from instances of the source meta-model.

\subsection{Model Validation}

While source and target meta-models impose {\it structural constraints (SC)} on {\it source} and {\it target models},
we can specify several {\it semantic requirements (SR)} on source and target models. In addition, we
can describe cross requirements on models. In Table \ref{requirements}
we can see a set of requirements classified as (SC) and (SR).  
Some requirements express conditions on {\it well-formed models} (WF), while some of them are required
by the transformation, that is, they are {\it transformation requirements} (TR).

\begin{table}
{\scriptsize

\begin{tabular}{|ll|}
\hline
{\bf Source Model} & \\
\hline
(1) All attributes of a data have distint names (SR) (WF) &
(2) Each data has a unique key attribute (SR) (TR) \\
(3) Each data has a key attribute (SR) (TR) &
(4) Each attribute is associated to exactly one data (SC) (WF)\\
(5) Each data is contained in exactly one store (SC) (WF) &
(6) All data have distinct names (SR) (TR)\\
(7) All data have distinct containers (SR) (TR) &
(8) Each qualifier is associated to exactly one role (SC) (TR)\\
(9) All qualifier names of a role are distinct (SR) (TR) &
(10) All qualifiers are key attributes (SR) (WF)\\
(11) Each relation has two roles (SC) (WF) &
(12) All relation names are distinct (SR) (WF) \\
(13) Each role is associated to exactly one relation (SC) (TR) &
(14) Each role is associated to exactly one data (SC) (TR) \\
(15) All role names of a data are distinct (SR) (TR) &
(16) Each store is associated to exactly one data (SC) (WF) \\
\hline
{\bf Target Model} &\\
\hline
(17) All col names of a row are distinct (SR) (WF) &
(18) All foreign names of a row are distinct (SR) (WF) \\
(19) All key names of a row are distinct (SR) (WF) &
(20) All foreigns of a row are keys of another row (SR) (WF)\\
(21) Each table is associated to exactly one row (SC) (WF) &
(22) Each row is associated to exactly one table (SC) (WF) \\
(23) Each key is associated to exactly one row (SC) (TR) &
(24) Each col is associated to exactly one row (SC)  (TR) \\
(25) Each foreign is associated to exactly one row (SC) (TR) &
(26) All table names are distinct (SR) (WF) \\
(27) All row names are distinct (SR) (WF) &
(28) All rows have exactly one key (SC) (TR) \\
(29) All rows have either all keys and cols or all foreigns (SR) (TR) &\\
\hline
{\bf Cross Requirements} &\\
\hline
(30) Key and col names and types are names and types of attributes &
(31) Table names are either container names or role names\\
(32) Row names are data names or concatenations of role and data names&
(33) Foreign names are concatenations of roles, data and keys\\
\hline
\end{tabular}
}
\caption{Model validation: requirements}\label{requirements}
\end{table}

\begin{itemize}
\item[--] For instance, (2) requires that each data has a unique key attribute. This is a semantic requirement. Key attributes
are attributes having {\it key} set to {\it true}, and the existence of a unique key attribute cannot be expressed
in the UML diagram. Moreover, this requirement is a pre-condition of the transformation because key attributes are used
as foreign keys in the target model. 
\item[--] Case (5) is an structural constraint on well-formed models: each data is associated
to exactly one store. It is not needed in the transformation and can be expressed in the UML diagram with a cardinality constraint.
\item[--] Cases (6), (7), (9), (12) and (15) are related to naming of elements of source models, and therefore they 
are semantic requirements.
They (except (12)) are required by the transformation: data and container names are used for naming tables and rows in the target model, while role and qualifier names (concatenated with data names) are also used for naming rows and foreign keys.
\item[--] (12) is required on a well-formed source model. 
\item[--] In the target model tables, rows, cols, keys and foreigns are not shared (cases (21)-(25)). 
\item[--] Case (20) is a semantic requirement that describes the relationship between foreign keys and
keys in a well-formed target model. 
\item[--] Case (29) is required by the transformation which assigns either keys and cols or foreigns
to rows. 
\item[--] Finally, cases (30)-(33) describe the relationship between names of the target model and names of the source model.
\end{itemize}

It is worth observing that the requirements about source and target models in isolation are not enough for the soundness
of the transformation. For instance, source and target models can both have keys, but a cross requirement is needed:
the keys of the target models are the keys of the source model. 

\section{Prolog for Model Transformation and Validation}

In this section, we will show how Prolog can be used for defining transformation and validation rules
in our approach. With this aim, we have to consider the 
following elements.

\subsection{Prolog based Transformation}

The Prolog interpreter has to import and export OWL files.
This is the case of SWI-Prolog which includes a library to import and export RDF(S)/OWL triples.
The SWI-Prolog library stores RDF triples in a da\-ta\-ba\-se, and they can be retrieved with the predicate
{\tt rdf}. The RDF library includes predicates: 
\begin{itemize}
\item[--] {\tt rdf\_reset\_db/0} which resets the database,
\item[--] {\tt rdf\_lo\-ad(+File,+Options)} for importing triples, 
\item[--] {\tt rdf\_save(+File)} for exporting triples,
\item[--] and finally, {\tt rdf\_\-a\-ssert(+Su\-b\-ject,+Pro\-per\-ty,+Ob\-ject)} for inserting a new triple in the current database.
\end{itemize}
A Prolog predicate {\tt trans\-form(+Sour\-ce\-Mo\-del\-Fi\-le,+Tar\-get\-Mo\-del\-Fi\-le)} has been defined to transform
a source model (stored in a OWL file) into a target model (stored also in a OWL file). The Prolog code of such 
predicate is as follows:\\

\begin{lstlisting}
transform(_,_):-rdf_reset_db,fail.
transform(_,_):-retractall(new(_,_,_)),fail.
transform(FileIn,_):-rdf_load(FileIn,[]),fail.
transform(_,_):-newrdf(A,B,C),assert(new(A,B,C)),fail.
transform(_,_):-rdf_reset_db,fail.
transform(_,_):-new(A,B,C),rdf_global_term(B,D),rdf_assert(A,D,C),fail.
transform(_,FileOut):-rdf_save(FileOut),rdf_reset_db.
\end{lstlisting}

The transformation rules define new triples representing the target model. Hence, a new predicate called
{\tt newrdf} is defined by the transformation rules.
For instance, the following rules define the individuals  of the class {\it table} of the model B
from the model A of the case study:\\

\begin{lstlisting}
newrdf(IdTable,rdf:type,'http://metamodelB.ecore#table'):-
					rdf(IdData,rdf:type,'http://metamodelA.ecore#data'),
					generate_id([IdData,'table1'],IdTable).
					
newrdf(IdTable,rdf:type,'http://metamodelB.ecore#table'):-   
					rdf(IdRole,'http://metamodelA.ecore#role.navigable',E),
					E=literal(type(_,true)),
					generate_id([IdRole,'table2'],IdTable).
\end{lstlisting}

The first rule defines triples {\scriptsize\tt (IdTable,rdf:type,'http://me\-ta\-mo\-del\-B.\-e\-co\-re\#ta\-ble')} obtained
from triples {\scriptsize\tt (IdData,rdf:type,'http://\-me\-ta\-mo\-del\-A.e\-co\-re\#data')}, where {\scriptsize\tt IdTable} is  the  identifier of the table, which is generated by the call {\scriptsize\tt generate\_id} from the data identifier 
{\scriptsize\tt IdData} and {\scriptsize\tt 'table1'}. 

The second rule defines the individuals of class {\it table} obtained
from navigable roles, which are generated from the role identifier {\scriptsize\tt IdRole}
and  {\scriptsize\tt 'table2'}. 

\noindent In such a way that the following Prolog goal obtains the tables of the target model:\\

\begin{lstlisting}
?- newrdf(IdTable,rdf:type,'http://metamodelB.ecore\#table').
    IdTable = 'http://metamodelA.ecore#02_Student_datatable1' ;
    IdTable = 'http://metamodelA.ecore#09_Course_datatable1' ;
    IdTable = 'http://metamodelA.ecore#13_register_roletable2' ;
false.
\end{lstlisting}

which represent the individuals of classes {\it Student}, {\it Course} and {\it register} of Figure \ref{fig:metamodelObjectB}.
Now, the individuals of the class {\it row} of Figure \ref{fig:metamodelObjectB} can be defined as follows:\\

\begin{lstlisting}
newrdf(IdRow,rdf:type,'http://metamodelB.ecore#row'):-
					rdf(IdData,rdf:type,'http://metamodelA.ecore#data'),
					generate_id([IdData,'row1'],IdRow).
					
newrdf(IdRow,rdf:type,'http://metamodelB.ecore#row'):-   
					rdf(IdData,'http://metamodelA.ecore#data.role_of',IdRole),
					rdf(IdRole,'http://metamodelA.ecore#role.navigable',E),
					E=literal(type(_,true)),
					generate_id([IdRole,IdData,'row2'],IdRow).
\end{lstlisting}

The first rule defines the individuals of the class {\it row} obtained from instances of data (i.e., the identifiers of
{\it the\_courses} and {\it the\_students}), and the
second rule defines the individuals of the class {\it row} 
obtained from navigable data roles (i.e., the identifier of {\it registerCourse}).

Now, {\it key}, {\it col} and {\it foreign} elements have to be defined. For instance,
the individuals of the class {\it foreign} are defined as follows:\\

\begin{lstlisting}
newrdf(IdForeign,rdf:type,'http://metamodelB.ecore#foreign'):- 
						rdf(IdRole,'http://metamodelA.ecore#role.navigable',E),
						E=literal(type(_,true)),
					 	rdf(IdRole,'http://metamodelA.ecore#role.is',IdQualifier),
						rdf(IdData,'http://metamodelA.ecore#data.role_of',IdRole),
						generate_id([IdRole,IdData,IdQualifier,'foreign1'],IdForeign).
						 

newrdf(IdForeign,rdf:type,'http://metamodelB.ecore#foreign'):- 					 
						rdf(IdRole,'http://metamodelA.ecore#role.navigable',E),
						E=literal(type(_,true)),
					 	rdf(IdRole,'http://metamodelA.ecore#role.has_role',IdRelation),
						rdf(IdRelation,'http://metamodelA.ecore#relation.is_role',IdRole2),
						rdf(IdRole2,'http://metamodelA.ecore#role.is',IdQualifier),
						IdRole2\==IdRole,
						rdf(IdData,'http://metamodelA.ecore#data.role_of',IdRole),
						generate_id([IdRole2,IdData,IdQualifier,'foreign2'],N).
\end{lstlisting}

In this case, instances of the class {\it foreign} are obtained from navigable roles, using the identifier of the qualifier
and the identifier of the role to generate the identifier. Now, the association roles of the Figure \ref{fig:metamodelObjectA} have to be defined. For instance, the role {\it has}  from the class {\it table} of Figure \ref{fig:metamodel}
is defined as follows:\\

\begin{lstlisting}
newrdf(IdTable,'http://metamodelB.ecore#table.has',IdRow):-
			rdf(IdData,rdf:type,'http://metamodelA.ecore#data'),
			generate_id([IdData,'table1'],IdTable),
			generate_id([IdData,'row1'],IdRow).
			 
newrdf(IdTable,'http://metamodelB.ecore#table.has',IdRow):- 
				rdf(IdData,'http://metamodelA.ecore#data.role_of',IdRole),
				rdf(IdRolw,'http://metamodelA.ecore#role.navigable',E),
				E=literal(type(_,true)),
				generate_id([IdRole,'table2'],IdTable),
				generate_id([IdRole,IdData,'row2'],IdRow).
\end{lstlisting}

The first rule defines the rows of the tables obtained from instances of {\it data}, and the second rule defines
the rows of the tables obtained from navigable roles. 

Finally, attributes of the classes of the target metamodel of  Figure \ref{fig:metamodel}
have to be defined. For instance, {\it name} of class {\it table} is defined 
as follows:\\

\begin{lstlisting}
newrdf(IdTable,'http://metamodelB.ecore#table.name',Name):-
						rdf(IdData,'http://metamodelA.ecore#data.container',Name),
						generate_id([IdData,'table1'],IdTable).

newrdf(IdTable,'http://metamodelB.ecore#table.name',Name):-			 
					rdf(IdRole,'http://metamodelA.ecore#role.name',Name),
					rdf(IdRole,'http://metamodelA.ecore#role.navigable',C),
					C=literal(type(_,true)),
					generate_id([IdRole,'table2'],IdTable).
\end{lstlisting}

\noindent where the table names are obtained from container names (i.e., {\it the\_students} and {\it the\_courses}). 
 
\subsection{Prolog based Validation}

Model validation is achieved with Prolog. Table \ref{rulesvalidation}
includes some of the Prolog rules of the requirements expressed in Table \ref{requirements}. The full set of rules
can be downloaded from \url{http://indalog.ual.es/mdd}. 

\begin{table}[!ht]
{\scriptsize
\begin{lstlisting}
(1) attribute_distinct_names:-
				       rdf(Data,mmA:'data.attr_of',Att1),rdf(Data,mmA:'data.attr_of',Att2),Att1\=Att2,
				       rdf(Att1,mmA:'attribute.name',Name1),rdf(Att2,mmA:'attribute.name',Name2),
				       Name1=Name2.
(2) exists_key:- 
			         setof(Att,(rdf(_,mmA:'data.attr_of',Att),
				       rdf(Att,mmA:'attribute.key',literal(type(_,true)))),Keys),
			         Keys=[].
(3) unique_key :- 
			         rdf(Data,mmA:'data.attr_of',Att1),rdf(Data,mmA:'data.attr_of',Att2),Att1\=Att2,
			         rdf(Att1,mmA:'attribute.key',literal(type(_,true))),
			         rdf(Att2,mmA:'attribute.key',literal(type(_,true))).
(5) unique_store_data:- 
				       setof(Store,rdf(_,mmA:'data.contained_in',Store),Stores),Stores=[_,_|_].
(10) qualifiers_are_keys :- 
				       rdf(_,mmA:'qualifier.name',Name),
				       \+(rdf(Attribute,mmA:'attribute.name',Name),
				      		rdf(Attribute,mmA:'attribute.key',literal(type(_,true)))).
(11) two_roles_relation :- 
				       setof(Role,rdf(_,mmA:'relation.is_role',Role),Roles),Roles\=[_,_].
(17) unique_col_names_row :-
					     rdf(Row,mmB:'row.is_col',Col1),rdf(Row,mmB:'row.is_col',Col2),Col1\=Col2,
					     rdf(Col1,mmB:'col:name',Name1),rdf(Col2,mmB:'col:name',Name2),Name1=Name2.
(20) foreign_keys:-
			         rdf(Row,mmB:'row.is_foreign',Foreign),
			         rdf(Row,mmB:'row.name',NRow),NRow=literal(type(_,RN)),
			         rdf(Foreign,mmB:'foreign.name',NFor),NFor=literal(type(_,FN)),
			         concat(RN,NKey,FN),
			         \+rdf(_,mmB:'key.name',literal(type(_,NKey))).
(21) unique_table_row :- 
				       rdf(Table,mmB:'table.has',Row1),rdf(Table,mmB:'table.has',Row2),Row1\=Row2.
(29) well_formed_rows:-
				       rdf(Row,mmB:'row.is_key',_),rdf(Row,mmB:'row.is_foreign',_).
     well_formed_rows:-
				       rdf(Row,mmB:'row.is_col',_),rdf(Row,mmB:'row.is_foreign',_).				
(31) containers_or_roles :- 
					     rdf(_,mmB:'table.name',Name),
					     \+rdf(_,mmA:'role.name',Name),\+rdf(_,mmA:'data.container',Name).
(32) data_or_roles_and_data :- 
					     rdf(_,mmB:'row.name',Name),Name=literal(type(_,N)),
					     \+rdf(_,mmA:'data.name',Name), 
					     \+(rdf(_,mmA:'role.name',Name1),rdf(_,mmA:'data.name',Name2),
						         Name1=literal(type(_,N1)),Name2=literal(type(_,N2)),concat(N1,N2,N)).
\end{lstlisting} 
}
\caption{Model validation: requirements}\label{rulesvalidation}

\end{table}

For validating the requirements on models, we can call the rules and in the case of success it indicates that the requirement is violated. In other words, the condition of the rule expresses the negation of the requirement.  Prolog meta programming predicates are used. For instance, case (2) uses the {\tt\scriptsize setof} predicate
to collect the set of keys of a given data.

 \section{Related Work}

Validation and verification of model transformations is an emerging topic of research. 
We have found some similarities of our approach with the work proposed in \cite{atlvalidation}. The authors
work in the context of the ATLAS Transformation language (ATL) and OCL, but handle the same kind of
properties of our approach (unique names for relations and attributes together with existence of keys).

A more general framework for transformation validation and verification is proposed
in \cite{cabot} including verification and validation of properties about transformation rules.
Our approach focused on properties about meta-models, assuming that when some requirement
is violated either source models or rules are incorrect.  

Prolog has been also used in the {\it Model Manipulation Tool (MoMaT)} \cite{Harald} for representing and verifying models. In \cite{khai2011prolog} they propose consistency checking of class and sequence diagrams based on Prolog. Consistency checking rules as well as UML models are represented by Prolog, and a Prolog reasoning engine is used to automatically find inconsistencies.

On the other hand, logic programming based languages have already 
been explored in the context of model engineering in some works. 

A first approach is \cite{Missing}, which describes the attempts to adopt several technologies for model transformation including
logic programming. Particulary, they focused on \emph{Mercury} and \emph{F-Logic} logic languages. 
The approach \cite{ILP2} has  introduced {\it inductive logic programming} in model transformation. The motivation of the work is that designers need to understand how to map source models into target models. With this aim, they are able to derive transformation rules from an initial and critical set of elements of the source and target models. The rules are generated in a (semi-) automatic way.  
 
The \emph{Tefkat} language \cite{Tefkat1,Tefkat2} is a declarative language whose syntax resembles a logic language with some differences (for instance, it incorporates a {\it forall} construct 
for traversing models). In this framework, in \cite{Deltaware}, they propose
metamodel transformations in which evolutionary aspects are formalised using the Tefkat language.
 
In \cite{Goldberg}, they present a declarative approach for modeling requirements (designs and patterns)
which are encoded as Prolog predicates. A search routine based on Prolog returns program fragments
of the model implementation. Traceability and code generation are based on logic programming. They use
{\it JTransformer}, which is a logic-based query and transformation engine for Java code, based on the Eclipse IDE.

Logic programming based model querying is studied in \cite{Dohrmann}, in which logic-based facts represent meta-models. 
In \cite{Schatz} they study a transformation mechanism for the EMF Ecore platform using Prolog as rule-based engine.
Prolog terms are used to represent models and predicates are used for deconstructing and reconstructing a term of a model. 

{\it Abductive logic programming} is used  in \cite{hettel2009towards} 
for {\it reversible} model transformations, in which changes of the source model
are computed from a given change of the target model. 
Finally, in \cite{Opoka}, they have compared OCL and Prolog for querying UML models. They have found that Prolog is faster
when execution time of queries is linear.

 \section{Conclusions and Future Work}
 
In this paper we have presented a framework for the specification and validation of model transformations 
with Prolog rules, using the representation of UML models
by ODM.  Our approach has been applied to a well-known example of model transformation in which an UML class diagram representing a ER diagram is transformed
into a UML diagram representing a relational database.  We have validated our proposal 
with a prototype developed under SWI-Prolog. 

Our approach has to be extended in the future as follows:
\begin{itemize}
\item[--] Firstly, we would like to improve our prototype. Particularly, validation is now achieved by Prolog rules
in which success and fail is returned. We would like to show more detailed 
analysis results,  showing the model elements that violate the requirements, justifications, diagnosis, reparations, etc.
\item[--] Secondly, we would like to test our approach with other  
UML diagrams and transformations, and also with bigger examples; 
\item[--] Thirdly, we are also interested in the use of our approach for model driven development 
of user interfaces in the line of our previous works \cite{luisjvlc,luiscomp2};
\item[--] Finally, we believe that our work will lead to the development of a logic based tool for transformation
and validation of models.
\end{itemize}

\bibliographystyle{eptcs}
\bibliography{biblio}
\end{document}